\documentclass[aps,prd,a4paper,groupedaddress,twocolumn,preprintnumbers,floatfix,nofootinbib]{revtex4}
\usepackage[english]{babel}
\usepackage{amsmath}
\usepackage{graphicx}
\usepackage{color}
\usepackage{amssymb}
\usepackage{hyperref}
\usepackage{dcolumn}
\usepackage{bm}
\begin{document}
\preprint{DCP-11-02}

\title{New Higgs signals from a multi-scalar model with flavor symmetry}

\author{
Alfredo Aranda,$^{1,2,3}$\footnote{Electronic address:fefo@ucol.mx} 
J. Lorenzo D\'iaz-Cruz,\footnote{Electronic address:jldiaz@fcfm.buap.mx} $^{3,4}$ Alfonso Rosado$^{3,5}$\footnote{Electronic address:rosado@ifuap.buap.mx}}

\affiliation{$^1$Facultad de Ciencias - CUICBAS,\\ 
  Universidad de Colima, Colima, M\'exico\\
  $^2$ Abdus Salam - ICTP, Trieste, Italy \\
  $^3$Dual C-P Institute of High Energy Physics, M\'exico \\
  $^4$C.A. de Particulas, Campos y Relatividad,
  FCFM-BUAP, Puebla, Pue., Mexico \\
  $^5$Instituto de F\'isica, BUAP, Puebla, Pue., Mexico}

\date{\today}

\begin{abstract}
A scenario is presented where the $s$, $c$, and $b$ quark fusion  Higgs production cross sections are enhanced with respect to those of the Standard Model. In particular the $c$ quark fusion production is very important and can account for a significant contribution at the Large Hadron Collider. The light Higgs couplings to vector bosons are sufficiently suppressed to allow its mass to lie below the LEP bound of $115$~GeV and due to enhanced couplings to second family fermions, the Higgs decay to $\mu$ pairs is large enough to be detectable. This is accomplished with a model incorporating three Higgs doublets charged under a flavor symmetry.
\end{abstract}
\maketitle

Electroweak symmetry breaking in the Standard Model is achieved by the Higgs mechanism, where the presence of an SU(2)$_W$ doublet scalar field - the Higgs field - with a non zero vacuum expectation (vev) leads to the breaking of SU(2)$_W \times$U(1)$_Y$ down to U(1)$_{em}$, rendering three massive gauge bosons ($Z$ and $W^{\pm}$), the massless photon, and a not yet observed massive scalar. Furthermore, the coupling of the Higgs field to fermions also generates their mass terms. 

Physics beyond the Standard Model tries to explain what the Standard Model leaves unexplained. One example is the exuberant distribution of fermion mass scales (including neutrino masses) and their mixing angles. Other interesting questions are related to the number of generations (three in the Standard Model), the hierarchy of energy scales, the apparently accidental global symmetries present in the Standard Model such as those associated to Baryon and Lepton number, etc. 

Of course there has also been an intensive experimental program dedicated to the search of deviations from the predictions of the Standard Model. With the salient exception of neutrinos oscillations, which however can in principle be easily incorporated to the Standard Model, there has been no deviation. This is a remarkable situation since any Physics Beyond the Standard Model proposal necessarily involves the prediction of deviations and their no observation puts stringent constraints to those approaches. Of present relevance is the fact that the Large Hadron Collider (LHC) has already provided data that is pushing the most popular candidate, namely the Minimal Supersymmetric Standard Model, 
to remote corners of parameter 
space~\cite{Khachatryan:2010uf,Khachatryan:2010mp,Aad:2011hh,Aad:2011xm,Chatrchyan:2011bj}, and it will certainly bring very interesting results within perhaps a couple of years.

Given this situation it is interesting to explore what kind of minimal additions to the Standard Model one can envisage such that, without explaining all of the open problems, one can still obtain interesting non-canonical phenomenology explorable at the LHC. 

This letter investigates the possibility of having three Higgs doublets and a flavor symmetry in such a way that the Higgs fields transform non-trivially under it. This is not a new idea~\cite{Aranda:2000zf,DiazCruz:2002er} and in fact most models of flavor symmetries, specially the so called renormalizable models (see for 
example~\cite{renormalizable}), include several Higgs doublets (and sometimes triplets) with non trivial flavor properties. The idea here is to explore the simplest case where there is only one Higgs doublet per generation and determine its phenomenology. It turns out that it is possible to obtain a scenario where, for example: the lightest scalar couplings to $W$ and $Z$ are suppressed, which allow its mass to be lower than the $115$~GeV LEP bound; the coupling of the lightest scalar to the top quark is significantly suppressed and turns down the usual gluon fusion production mechanism for the Higgs; the Higgs production cross section through charm and strange fusion at the LHC can be $2 - 3$ orders of magnitude larger than that one of the SM. 

To see how this works consider a model with the Standard Model fermion content plus three Higgs SU(2) doublets. Assume there is a softly broken global U(1) flavor symmetry under which the fields in the model are charged as:~\footnote{This model is presented as an existence proof to show the scheme. A more detailed treatment with specific models based on discrete flavor symmetries that include the neutrino sector will be presented in a separate publication.}
\begin{center}
\begin{tabular}{||c|c|c|c|c|c|c|c|c||c|c|c||}
\hline
$Q^0_1$ & $Q^0_2$ & $Q^0_3$ & $u^0_1$ & $u^0_2$ & $u^0_3$& $d^0_1$ & $d^0_2$ & $d^0_3$& $\Phi_1$ & $\Phi_2$ & $\Phi_3$\\
\hline
$-a$ & $-b$ & $0$ & $a$ & $b$ & $0$& $-2b$ & $-a-b$ & $-b$ & $a+b$ & $b$ & $0$\\
\hline
$L^0_1$ & $L^0_2$ & $L^0_3$ &  $\bullet$& $\bullet$ & $\bullet$& $e^0_1$ & $e^0_2$ & $e^0_3$& $\bullet$ & $\bullet$ & $\bullet$ \\
\hline
\end{tabular}
\end{center}
where the charges are chosen so that $|a| \neq \{|b/2|, |b|, |2b|\}$, and where the fermion fields are given in the weak basis.

With these assignments the allowed Yukawa terms are given by (for quarks)
\begin{eqnarray}\label{yukawas} \nonumber
{\cal L}_Y &=& [y^u]_{1,2} \overline{Q}^0_1 \tilde{\Phi}_1 u^0_{R2} + [y^u]_{2,1} \overline{Q}^0_2 \tilde{\Phi}_1 u^0_{R1} \\ \nonumber
&+& [y^u]_{2,3} \overline{Q}^0_2 \tilde{\Phi}_2 u^0_{R3} + [y^u]_{3,2} \overline{Q}^0_3 \tilde{\Phi}_2 u^0_{R2} \\ \nonumber 
&+& [y^u]_{3,3} \overline{Q}^0_3\tilde{\Phi}_3 u^0_{R3} + [y^d]_{1,2} \overline{Q}^0_1 \Phi_2 d^0_{R2}  \\ \nonumber 
&+& [y^d]_{2,1} \overline{Q}^0_2 \Phi_2 d^0_{R1} + [y^d]_{2,3} \overline{Q}^0_2 \Phi_3 d^0_{R3} \\
&+& [y^d]_{3,2} \overline{Q}^0_3 \Phi_1 d^0_{R2} + [y^d]_{3,3} \overline{Q}^0_3 \Phi_2 d^0_{R3} + h.c. \ ,
\end{eqnarray}
where $y^{u}$ and $y^d$ denote the Yukawa matrices in the up and down sectors respectively, 
$\tilde{\Phi} \equiv i\sigma_2 \Phi^*$, and where the charged lepton Yukawa terms are similar to the down-type quark terms.

Spontaneous breaking of electroweak symmetry is triggered when $\Phi_1$, $\Phi_2$ and $\Phi_3$ acquire vevs ($v_1$, $v_2$, and $v_3$, in obvious notation,  and where we assume CP-even vevs). Denoting these fields by $\phi^0_a= \frac{1}{\sqrt{2}} ( v_a+ \phi^0_{Ra}+ i \phi^0_{Ia})$ Eq.~(\ref{yukawas}) then leads
the Yukawa quark mass matrices
\begin{eqnarray}\label{qyukawamatrices}
M^0_u &=&
\left(
	\begin{array}{ccc}
	0 & v_1 y^u_{12} & 0 \\
	v_1 y^u_{21} & 0 & v_2 y^u_{23} \\
	0 & v_2 y^u_{32}& v_3 y^u_{33}
	\end{array}
\right) \\
M^0_d &=&
\left(
	\begin{array}{ccc}
	0 & v_2 y^d_{12} & 0 \\
	v_2 y^d_{21} & 0 & v_3 y^d_{23}\\
	0 & v_1 y^d_{32} & v_2 y^d_{33}
	\end{array}
\right) \ 
\end{eqnarray}
(the charged lepton mass matrix is obtained from $M_d^0$ replacing $y^d_{ij}$ with $y^l_{ij}$).

Note that the general form of the mass matrices is given by the Fritzsch-like texture~\cite{Fritzsch1} ($f=u,d,l$)
\begin{eqnarray}
M^0_f =
\left(
	\begin{array}{ccc}
	0 & A_f  & 0 \\
	A'_f & 0 & B_f \\
	0 & B'_f & C_f
	\end{array}
\right) 
\end{eqnarray}
where the difference from the original Fritzsch texture is that they deviate from the Hermitian Nearest-Neighbor Interaction form (NNI). As discussed in \cite{Branco:2010tx}, only small deviations from hermiticity are required in order to fit the CKM matrix, and one can perform a perturbative diagonalization by splitting the mass matrix in the following way:
\begin{eqnarray}
M^0_f = H^0_f + \Delta M_f \ .
\end{eqnarray}

The hermitian mass matrix $H^0_{f}$ is obtained from $M^0_f$ by taking $y^f_{21}=y^{f*}_{12}$ and 
$y_{32}=y^{*}_{23}$, while $\Delta M$ parametrizes the deviation from hermiticity. 
The diagonal mass matrix is obtained by the usual bi-unitary transformation 
$\bar{M}_f = O^f_L M^0_f O^{f\dagger}_R$ whereas for the hermetic mass matrix 
a single matrix $O_L=O_R= O_u$ suffices. Thus $\bar{H}_{f} = O_f H^0_{f} O^{\dagger}_f$. 
and the matrices  $O_f$ can be expressed in terms of mass eigenvalues. 
The diagonalization matrices for the complete mass matrices $M^0_f$ are then given by 
$O^f_L= O_f(1+X_f)$ and $O^f_R= O_f(1-X_f)$, where $X_f$ is obtained perturbatively~\cite{Fritzsch2}.

The fermion interactions, expressed in the mass basis, with the neutral scalar fields are given by
\begin{eqnarray}
 {\cal{L}}_{Yf} =  \bar{f}_{Li} (O_L)_{ik} (Y^f_a)_{kl} (O^{\dagger}_R)_{lj}  f_{Rj} \phi^0_a +h.c.
\end{eqnarray}
where the matrices $Y^u_a$ are obtained from the $y^u$ in Eq.~(\ref{yukawas}) for each $\Phi_a$.

Denoting the neutral CP-even Higgs mass eigenstates by $h_a$, where $\phi^0_{Ra} = U_{ab} h_b$, with $U$ the unitary matrix that diagonalizes the scalar squared mass matrix, one obtains the following Yukawa interactions:
\begin{eqnarray}
{\cal{L}}_{Yf} &=& \frac{1}{2} \bar{f}_{i} \left( [ (\Lambda^f_b)_{ij} +  (\Lambda^{f*}_b)_{ji} ] \right. \nonumber \\ 
&+& \left. [ (\Lambda^f_b)_{ij} - (\Lambda^{f*}_b)_{ji} ] \gamma_5 \right)  f_{j} h^0_b + h.c.,
\end{eqnarray}
where
\begin{eqnarray}
 (\Lambda^f_a)_{ij} = (O_L)_{ik} (Y^f_b)_{kl} (O^{\dagger}_R)_{lj}  U_{ab} \ .
\end{eqnarray}

Neglecting first generation masses (a sensible approximation for the phenomenology of interest)
the matrices $O_{L,R}$ for up-type quarks can be written as
\begin{eqnarray}
O_{L,R} =
\left(
	\begin{array}{ccc}
	1 & 0  & 0 \\
	0 &  c (1+r \epsilon) & s(1-\epsilon) \\
	0 & -s (1-\epsilon)  & c (1+r \epsilon)
	\end{array}
\right) 
\end{eqnarray}
where $c$ $(s)$ denote the cosine (sine) of the mixing angle in the up sector, $r= m_c/m_t$, and $\epsilon= (B_u-B'_u)/abs(B_u)$. The resulting coupling matrices take the form
\begin{eqnarray}\label{lambdaua}
\Lambda^u_a =
\left(
     \begin{array}{ccc}
     1 & 0  & 0 \\
     0 &  m_c [2 \frac{U_{2a}}{v_2} + \frac{U_{3a}}{v_3}]  &  \bar{m}_{ct} [\frac{U_{2a}}{v_2} + \frac{U_{3a}}{v_3}]  \\
     0 &  \bar{m}_{ct} [-\frac{U_{2a}}{v_2} + \frac{U_{3a}}{v_3}]   &  - 2m_c \frac{U_{2a}}{v_2} + m_t \frac{U_{3a}}{v_3}] 
	\end{array}\right) \ \
\end{eqnarray}
where $\bar{m}_{ct} = \sqrt{m_c m_t}$. Similar expressions are obtained for d-type quarks
and charged leptons.

Finally, the diagonal Higgs-fermion couplings $h_a f_i \bar{f}_i$, for all types of
quarks and leptons can be written as (no sum over $k$)
\begin{eqnarray}
 (\Lambda^f_a)_{kk} = g^{k}_{sm} \chi^{f_k}_{a}  = \frac{m_{k}}{v} \chi^{f_k}_{a} \ ,
\end{eqnarray}
where the couplings $\chi^{u_k}_{a}$ for up-type quarks can be easily read off Eq.~(\ref{lambdaua})
(and similarly for d-type quarks and charged leptons) and $v=246$~GeV. The resulting coefficients $\chi^{f_k}_{a}$ are given by
\begin{eqnarray}
 \chi^{s}_{a} &=& U_{1a} \frac{v}{v_1} \nonumber \ , \\
 \chi^{b}_{a} &=&  U_{2a} \frac{v}{v_2} - U_{1a} \frac{v m_s}{v_1 m_b} \nonumber \ , \\
 \chi^{c}_{a} &=& 2U_{2a} \frac{v}{v_2}+  U_{3a} \frac{v}{v_3} \nonumber \ , \\
 \chi^{t}_{a} &=&  U_{3a} \frac{v}{v_3} - 2 U_{2a} \frac{v m_c}{v_2 m_t} \nonumber \ , \\
 \chi^{\mu}_{a} &=& U_{1a} \frac{v}{v_1} \nonumber \ , \\
 \chi^{\tau}_{a} &=&  U_{2a} \frac{v}{v_2} - U_{1a} \frac{v m_{\mu}}{v_1 m_{\tau} } \ .
\end{eqnarray}

In order determine the matrix elements $U_{ab}$ one must analyze the scalar potential of the model, which in this case is given by
\begin{eqnarray} \label{potential} \nonumber
V_H &=& \mu_i^2 |\Phi_a|^2 + \lambda_i |\Phi_a|^4 
\\ \nonumber
&+&
\lambda_{ab; a<b}^{(1)} |\Phi_a|^2|\Phi_b|^2 
+ \lambda_{ab; a<b}^{(2)} \left|\Phi_a^{\dagger}\Phi_b\right|^2 
\\ \nonumber
&+&
\eta_{ab; 1<a\neq b}\left[ \left(\tilde{\Phi}_c^{\dagger} \Phi_b\right) \left(\Phi_b^{\dagger}\Phi_c \right)  + h.c.\right] \\
&+& f_{ab;a<b} \left(\Phi_a^{\dagger}\Phi_b + h.c.\right) \ ,
\end{eqnarray}
where all couplings are real and the terms proportional to $f_{ab}$ are the soft-breaking terms.

\begin{figure}[ht]
\includegraphics[width=6cm]{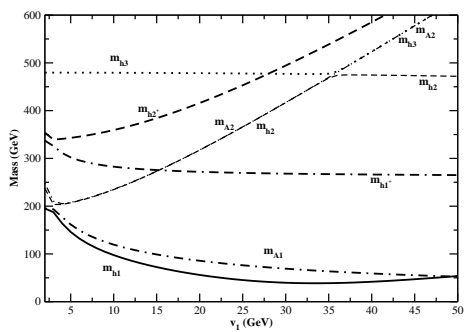}
\caption{Masses of the scalar particles in the model for the parameters given in the text with $v_1=5$~GeV.}
\label{fig:spectrum}
\end{figure}

A numerical analysis shows that $v_1 \simeq v_2$ must be small, which implies that $v_3 \sim$ EW-scale, and in order to obtain sizable masses for the pseudoscalars, $f_{ab}$ must be large. 
Taking O(1) values for all the other coefficients one can obtain acceptable scalar masses and vacuum stability (see figure~\ref{fig:spectrum}). A detailed analysis will be presented in a separate publication.

An interesting observation is that since $v_1 \simeq v_2 << v_3$, then the masses of the charm and bottom quarks are obtained with Yukawa couplings that can have the same size as that of the top quark Yukawa coupling, therefore leading to a large enhancement of their Higgs couplings (similarly for the muon and tau). This fact provides the possibility for a sizable (compared to the SM) production of Higgs bosons at LHC through s and c quark fusion. In table~\ref{table1} an example is presented where $v_1 = 7$~GeV and $v_2$ is varied from $3$ to $17$~GeV. The O(1) parameters and soft breaking masses from the scalar potential used in this example are given by $\lambda_1 = 1.5$, $\lambda_2 = 1.8$, $\lambda_3 = 1.9$, $\lambda^1_{12} = 0.7$, $\lambda^1_{13} = 0.5$, $\lambda^1_{23} = -1.2$, $\lambda^2_{12} = 2.1$, $\lambda^2_{13} = -2.2$, $\lambda^2_{23} = -2.5$, $\eta_{23}=-1.6$, $\eta_{32}=-1.9$, $f_{12}=-444$~GeV,  $f_{13}=-450$~GeV, and $f_{23}=-457$~GeV 

\begin{table}[ht]
\begin{tabular}{|c|| c|| c|c|c|c|c|c|c||}
\hline
$v_2$ & $m_{h_1}$ & $\chi^{c}_1$ & $\chi^{t}_1$ & $\chi^{s}_1$ & $\chi^{b}_1$  
&  $\chi^{\mu}_1$ & $\chi^{\tau}_1$ & $\chi^{W}_1$   \\
\hline
\hline
 $3$  & $121.4$ & $33.7$ & $-0.27$  & $34.4$ & $16.2$  & $34.4$  & $14.7$  & $0.05$ \\
\hline
 $5$  & $115.1$ & $41.1$ & $-0.33$  & $31.9$ & $19.9$  & $31.9$  & $18.6$  & $0.06$ \\
\hline 
 $10$ & $92.2$  & $38.5$ & $-0.29$  & $21.9$ & $18.8$  & $21.9$  & $17.9$ & $0.09$ \\
\hline  
 $17$ & $67.9$  & $25.7$ & $-0.15$  & $16.1$ & $12.5$  & $16.1$  & $11.8$ & $0.14$ \\
\hline
\end{tabular}
\caption{Values obtained for the couplings $\chi^{f_k}_a$ for $v_1 = 7$~GeV for the parameters in the scalar potential discussed in the text. Note that the value of $v_3$ is obtained from the relation $v_i^2 = (246 \ \rm{GeV})^2$. Dimension-full parameters are in GeV.}
\label{table1}
\end{table}

Note that the lightest Higgs coupling to vector boson pairs (W or Z) is very suppressed. This allows its mass to be lower than the $115$~GeV LEP bound. Also, its coupling to the top quark is significantly suppressed, which will make the usual gluon fusion mechanism for Higgs production smaller. On the other hand, the enhancement found for the Higgs couplings to the c,s and b quarks, will have a very interesting effect on the production of Higgs bosons at LHC through their fusion.

\begin{figure}
\includegraphics[width=7cm]{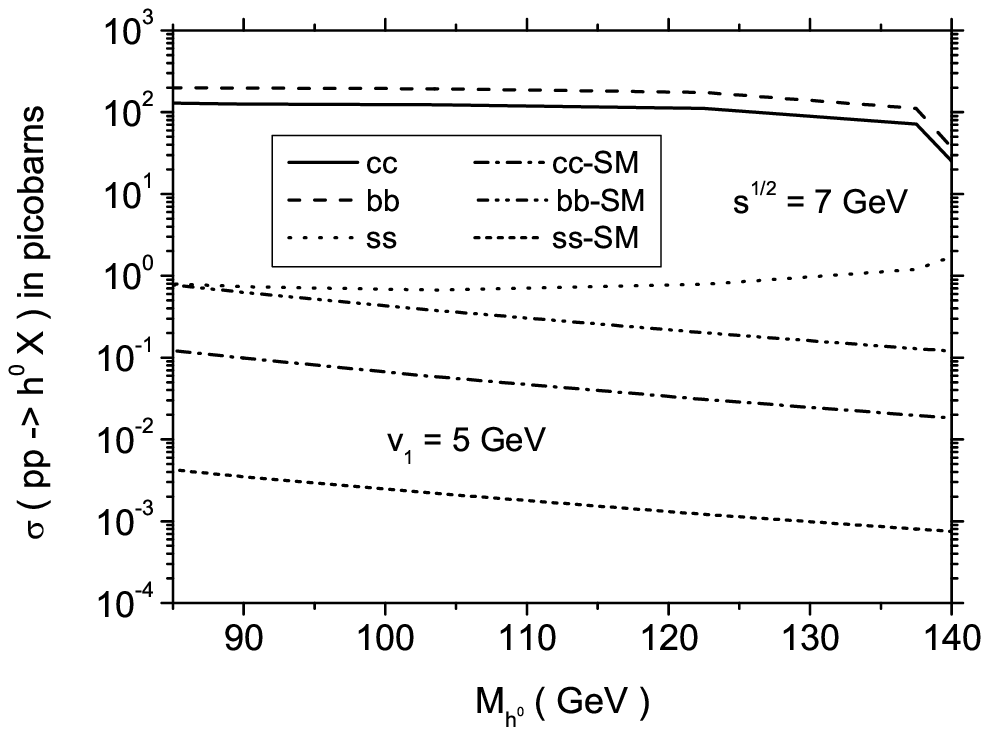} \\
\includegraphics[width=7cm]{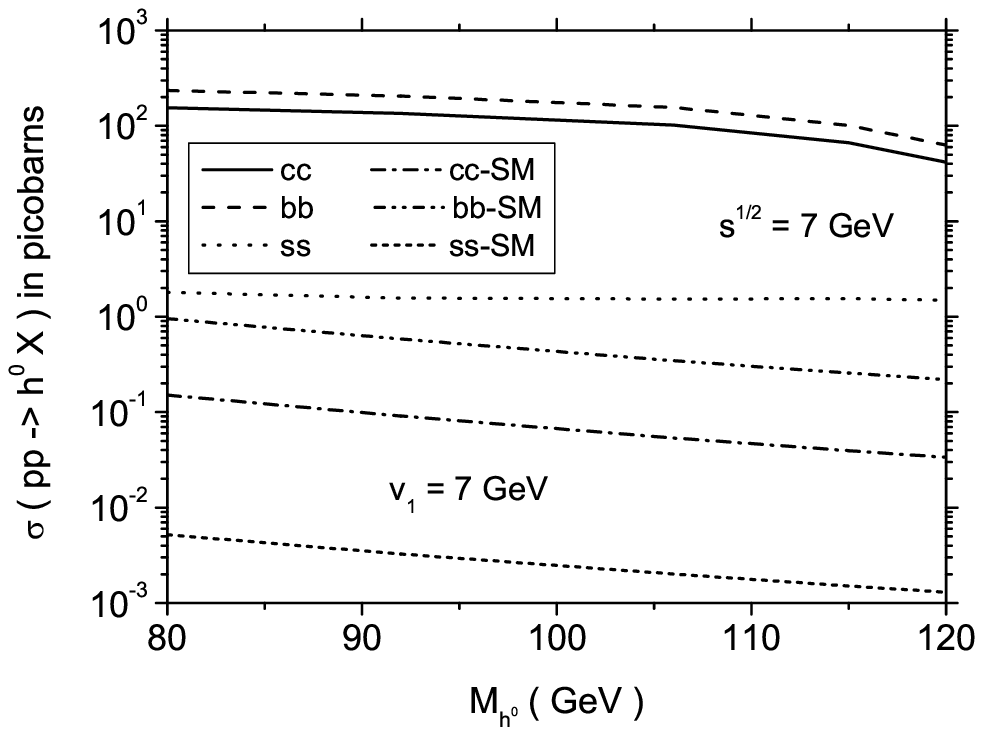}
\caption{Production cross sections for $s$, $c$, and $b$ quark fusion for the present model and those corresponding to the Standard Model. The upper plot is for a value of $v_1=5$~GeV while the lower one is for $v_1=7$~GeV. All other parameter correspond to the values quoted in the text. $h^0$ generically denotes the Standard Model Higgs field and the lightest field $h_1$.}
\label{fig:crossections}
\end{figure}

Using the CTEQ parton distribution functions~\cite{Pumplin:2002vw} one finds that
the cross section for the production of the light Higgs boson $h_1$, through charm and 
strange fusion at LHC7, is $2-3$ orders of magnitude larger than the SM result. This is shown in figure~\ref{fig:crossections} for two cases with $v_1=5$ and $7$~GeV respectively.

The decays of the Higgs boson are obtained to be:
\begin{eqnarray} 
BR(h_1 \to \mu^+ \mu^-)&=& 10^{-3} \\ 
BR(h_1 \to \tau^+ \tau^-)&=&5.3\times 10^{-2} \\
BR(h_1  \to \  c \ \ \bar{c} \ \ )&=& 0.2 \\
BR(h_1 \to \ s \ \ \bar{s} \ \ )&=&1.5\times 10^{-3} \\ 
BR(h_1 \to \ b \ \ \bar{b} \ \ )&=&0.74
\end{eqnarray}
Thus, with $10$~fb$^{-1}$ at LHC7, one gets about $1000$ events from charm fusion, followed by the decay $h_1 \to \mu^+ \mu^-$, this needs to be compared with the Drell-Yan background, which for a mass resolution of $5$~GeV, gives $1.21\times 10^{4} \pm 1.1\times 10^2$ events. Thus the signal of the model is clearly detectable.

A similar enhancement for the bottom quark and tau lepton Higgs coupling is found, 
which will also produce an enhancement of the production of the Higgs boson in association with b-
pairs at the LHC~\cite{Carena:1998gk}. In  this case one can have possible detectable signals coming from the
decays $h_1 \to b\bar{b}$, $h_1 \to \tau^+\tau^-$, $h_1 \to \mu^+\mu^-$, which will have final 
states consisting of $4b$ quarks, $2b$ quarks plus $2 \tau$ or $2b$ quarks plus $2 \mu$,
respectively (the signal to $\tau$ pairs is discussed in~\cite{Drees:1997sh}). As mentioned above a complete numerical analysis of these processes and the scalar sector will be presented in a separate publication but it can be noted that the enhancement for the Higgs-bottom coupling is well above the ones that could be tested at LHC14. As shown in~\cite{Balazs:1998nt,DiazCruz:1998qc}, one needs enhancements of size $2.4$ in oder to detect a Higgs particle with a mass of order $200$~GeV. This means that the signal in this model is clearly detectable at the LHC.

\begin{acknowledgments}
We thank Paolo Amore for reading the manuscript. This work was supported in part by CONACYT and SNI (Mexico). A.~A. thanks The Abdus Salam ICTP for its hospitality while part of this work was carried out.
\end{acknowledgments}

\end{document}